\documentclass{raa}
\usepackage{graphicx,times}
\usepackage{natbib}
\usepackage{amssymb,amsmath}
\usepackage[T1]{fontenc}
\usepackage{aecompl}
\bibpunct{(}{)}{;}{a}{}{,}
\usepackage[a4paper=true,dvipdfm=true,pagebackref=true]{hyperref}
\hypersetup{pdftitle = A New Strategy of Estimating Photometric Redshifts of Quasars, pdfauthor = Zhang Y. et al., pdfsubject= The subject, pdfkeywords = keyword1 keyword2 keyword3}
\hypersetup{colorlinks = true, linkcolor = green, anchorcolor = red, citecolor = blue, filecolor = red, pagecolor = red, urlcolor = red}

\begin{document}
\title{A New Strategy for Estimating Photometric Redshifts of Quasars$^*$ \footnotetext{\small $*$ Supported by the National Natural
Science Foundation of China.} }

 \volnopage{ {\bf 2019} Vol.\ {\bf X} No. {\bf XX}, 000--000}
   \setcounter{page}{1}

   \author{Yan-Xia Zhang\inst{1}, Jing-Yi Zhang\inst{1,2}, Xin Jin\inst{1,2}, Yong-Heng Zhao\inst{1}}
   \institute{ Key Laboratory of Optical Astronomy, National Astronomical Observatories,
Chinese Academy of Sciences, 20A Datun Road, Chaoyang District,
100101, Beijing, P.R.China; \\
        \and University of Chinese Academy of Sciences, Beijing 100049, China. {\it zyx@bao.ac.cn}\\
\vs \no
   {\small Received 2019 February; accepted 2019 June}
}

\abstract{Based on the SDSS and SDSS-WISE quasar datasets, we put forward two schemes to estimate the photometric redshifts of quasars. Our schemes are based on the idea that the samples are firstly classified into subsamples by a classifier and then photometric redshift estimation of different subsamples is performed by a regressor. Random Forest is adopted as the core algorithm of the classifiers, while Random Forest and $k$NN are applied as the key algorithms of regressors. The samples are divided into two subsamples and four subsamples depending on the redshift distribution. The performance based on different samples, different algorithms and different schemes are compared. The experimental results indicate that the accuracy of photometric redshift estimation for the two schemes generally improve to some extent compared to the original scheme in terms of the percents in $\frac{|\Delta z|}{1+z_{i}}<0.1$ and $\frac{|\Delta z|}{1+z_{i}}<0.2$ and mean absolute error. Only given the running speed, $k$NN shows its superiority to Random Forest. The performance of Random Forest is a little better than or comparable to that of $k$NN with the two datasets. The accuracy based on the SDSS-WISE sample outperforms that based on the SDSS sample no matter by $k$NN or by Random Forest. More information from more bands is considered and helpful to improve the accuracy of photometric redshift estimation. Evidently it can be found that our strategy to estimate photometric redshift is applicable and may be applied to other datasets or other kinds of objects. Only talking about the percent in $\frac{|\Delta z|}{1+z_{i}}<0.3$, there is still large room for further improvement in the photometric redshift estimation.
}

   \authorrunning{Y. Zhang et al. }            %author_head in even pages
   \titlerunning{A New Strategy}  % title_head in odd pages
   \maketitle

\section{INTRODUCTION}

With the development of large photometric survey projects (e.g. 2MASS, GALEX, the Sloan Digital Sky Survey (SDSS), Pan-STARRS, LSST), we face a photometric data deluge, which is the best test bed for various algorithms. Among them, the photometric redshift estimation is an important issue. Research on this aspect focuses on celestial objects, such as galaxies, quasars, supernovas, gamma-ray bursts and so on. The study of photometric redshifts is of great importance to the large scale structure of the Universe, the formation and evolution of galaxies, clustering of galaxies, distance measurement, and so on. There are lots of works on the photometric redshift measurement of distance objects including quasars, and especially galaxies. Furthermore, a large number of algorithms and tools on photometric redshift estimation are in development. The algorithms are grouped into two kinds: template-fitting and machine learning, for instance, Bayesian method \citep{ben00, ed06, Mortlock12}, color-redshift relation \citep{ric01, wu04, ball07}, $k$-Nearest Neighbors ($k$NN; \citealt{ball07, zhang13}), Gaussian process regression \citep{way06, way09, bon10}, sparse Gaussian process regression \citep{alm16a, alm16b}, Artificial Neural Networks (ANNs; \citealt{fir03, zhang09, yec10, cav12, bre13, cav17}), kernel regression \citep{wang07}, spectral connectivity analysis (SCA; \citealt{free09}), Random Forests (RFs) \citep{car10, sch17}, ArborZ \citep{ger10}, the extreme deconvolution technique \citep{bov12}, the Directional Neighborhood Fitting (DNF) algorithm \citep{de16}, a hybrid technique \citep{beck16}, Self-Organizing-Map (SOM; \citealt{way12, car14}), Clustering aided Back propagation Neural network (CuBANz; \citealt{sam17}), and Support Vector Machine (SVM; \citealt{jon17, sch17}).

To improve the accuracy of photometric redshift estimation,
researchers have considered new approaches or combined several methods. \citet{wolf09} combined $\chi^2$ template fits and empirical training sets into a single framework,
applied it to the SDSS Data Release 5 (DR5) quasars, and improved the accuracy of photometric redshift estimation.
\citet{lau11} put forward Weak Gated Experts (WGE) to derive photometric redshifts of galaxies and quasars through a combination of data mining techniques.
\citet{gor14} investigated different approaches and combined a template-fitting method and a neural network method for photometric redshifts of galaxies.
\citet{han16} integrated $k$NN and SVM for photometric redshift estimation of quasars.
\citet{hoy16} proposed Deep Neural Networks to estimate the photometric redshift of galaxies by using the full galaxy image in each measured band.
\citet{lei17} presented a new method for inferring photometric redshifts in deep galaxy and quasar surveys, which combines the
advantages of both machine learning methods and template fitting methods by building template spectral energy distributions (SEDs) directly from the spectroscopic training data.
\citet{wolf17} investigated the photometric redshift performance of several empirical and template methods, and kernel-density estimation (KDE) was the best for their case.
\citet{jou17} explored different techniques to reduce the photometric redshift outliers fraction with a comparison between the template fitting, neural
networks and RF methods. \citet{sp17a, sp17b} derived photometric redshifts using fuzzy archetypes and SOMs and
demonstrated that the statistical robustness and flexibility can be gained by combining template-fitting and machine-learning methods, and can provide useful insights into how astronomers may further exploit the color-redshift relation.
Since large numbers of images are available, it is applicable to directly use image data and save time by preprocessing image data.
\citet{dis18} probed deep learning to derive probabilistic photometric redshift directly from multi-band imaging data, rendering pre-classification of objects and feature extraction obsolete.

Although a large number of algorithms have been employed in this aspect, algorithms that perform well on galaxies may be not necessarily applicable for quasars. Because the accuracy of the photometric redshift estimation of quasars is not too satisfactory, there is still large room for improvement. Therefore, we have designed a new strategy to estimate the photometric redshifts of quasars.
The sample used for photometric redshift estimation is described in Section 2.
Then, the adopted methods are briefly introduced in Section 3. Based on the SDSS and SDSS-WISE samples, the different schemes of
photometric redshift estimation of quasars by $k$NN and RF are depicted in detail and compared in Section 4.
The discussion is presented in Section 5. Finally we summarize the results of this paper in Section 6.

\section{Sample}
The SDSS (\citealt{York00}) has been one of the most successful surveys in the history of astronomy. In particular, it has created the most detailed three-dimensional maps of the Universe ever complied, with deep multi-color images of one-third of the sky, and spectra for more than three million astronomical objects. We adopt the quasar sample from the Data Release 14 Quasar catalog (DR14Q) of SDSS-IV/eBOSS \citep{Paris18}. The DR14Q contains 526 356 unique quasars, of which 144 046 are new discoveries since the beginning of SDSS-IV. The catalog also includes previously spectroscopically-confirmed quasars from SDSS-I, II and III. Spectroscopic observations of quasars were performed over 9 376 deg$^2$ for SDSS-I/II/III and are available over 2 044 deg$^2$ for new SDSS-IV. Removing the records which contain default SDSS magnitudes, $z$Warning$=-1$ and full magnitude errors largeR than 5, the number of entries in the SDSS quasar sample reduces to 445 958. When further ruling out the records with default $W1$ and $W2$, the number of entries in the SDSS-WISE quasar sample amounts to 324 333. In this paper, we adopt AB magnitudes and convert the SDSS $u$-band and $z$-band magnitudes with $u_{\rm AB}=u'-0.04$ mag and $z_{\rm AB}=z'+0.02$. All magnitudes are corrected for Galactic extinction with the extinction values from DR14Q. The $W1$ (3.4$\mu$m) and $W2$ (4.6$\mu$m) of the Wide-field Infrared Survey Explorer (WISE; \citealt{Mainzer11}) are directly obtained from DR14Q and converted to AB magnitudes using $W1_{\rm AB}=W1+2.699$ and $W2_{\rm AB}+3.339$, and then extinction-corrected by the extinction coefficients $\alpha_{\rm W1},\alpha_{\rm W2}$=0.189, 0.146 with the extinction values from SDSS photometry. The AB magnitude conversion and extinction correction process are similar to the work of \citet{sch17}.

\section{Methods}

The $k$NN method belongs to the lazy learning family, which delays its learning until prediction. Its principle of operation is to find the $k$ training samples closest in distance to the new point, and predict the label from these. For the classification problem, the new point is labeled according to the majority of the $k$ closest neighbors. For applications involving regression, the prediction is the average of the $k$ closest neighbors. In general, the distance can be any metric measure, and standard Euclidean distance is the most common choice. To improve the query speed, a fast indexing structure such as a Ball Tree or KD Tree is adopted.

RF (\citealt{Breiman01}) is based on bagging models built using the Random Tree method, in which classification trees are grown on a random subset of descriptors (eg. \citealt{gao09}).
Each tree in the ensemble is constructed from a sample drawn with replacement (i.e., a bootstrap sample) from the training set. When splitting a node in the process of tree construction, the chosen split is no longer the best split among all of the features. However, this split is the best split among a random subset of the features. Based on this randomness, the bias of the forest usually slightly increases (with respect to the bias of a single non-random tree). However, due to averaging, its variance also decreases, usually more than compensating for the increase in bias, hence yielding an overall better model. Therefore RF uses the average to improve the predictive accuracy and control over-fitting.
Compared to \citet{Breiman01}, the scikit-learn implementation of RF combines classifiers by averaging their probabilistic prediction, instead of letting each classifier vote for a single class.

For these two methods, we use KNeighborsRegressor, RandomForestRegressor and RandomForestClassifier from the Python
module scikit-learn \citep{Ped11}.

\section{Photometric redshift estimation}

The redshift distribution of this sample is depicted in Figure~1. Because the quasar colors change with redshift and the dominating features appear in different bands with different redshifts, we divide the quasar sample into two classes and four classes according to the redshift range. The two classes are one with $0<z\le 2.2$ and the other with $2.2<z$. The four classes are ``vlowz" with $0<z \le1.5$, ``lowz" with $1.5<z\le 2.2$, ``midz" with $2.2<z\le 3.5$, and ``highz" with $3.5<z$ similar to \citet{sch17}. At the first break of $z=1.5$, the Lyman-alpha (Ly$\alpha$) emission line stays blueward of the $u$-band and the CIV emission line still remains in the $g$-band. Because the second break is at $z=2.2$, the Ly$\alpha$ emission line is just leaving the $u$-band. At $z=3.5$, a strong flux decreases in the $u$-band while the Ly$\alpha$ forest absorbs flux blueward of Ly$\alpha$ line. We apply these two classes and four classes to label the SDSS and SDSS-WISE quasar samples for the classification problem. In the following experiments, for the SDSS quasar sample, $r,u-g,g-r,r-i,i-z$ are taken as the input pattern, while for the SDSS-WISE quasar sample, $r,u-g,g-r,r-i,i-z, z-W1, W1-W2$ are adopted. The whole quasar samples from SDSS and SDSS-WISE are randomly separated into two-thirds for training and one-third for testing.

\begin{figure}
\centering
\includegraphics[bb=19 22 453 341,width=12cm]{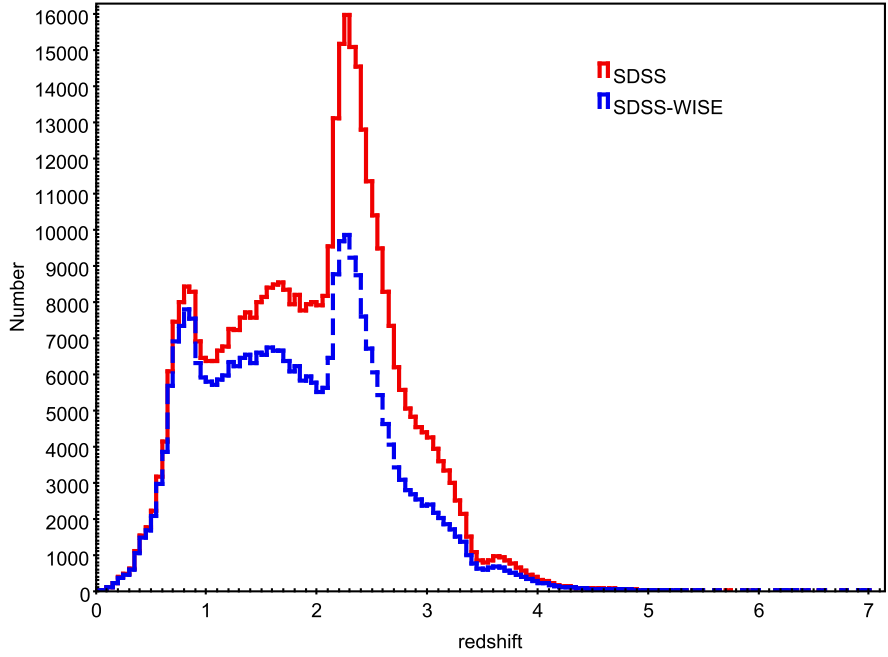}
\caption{Spectroscopic redshift distribution of the SDSS quasars (the red solid line histogram) and the SDSS-WISE quasars (the blue dashed line histogram). The bin size is $\Delta z=0.05$.}
\label{fig1}
\end{figure}

The problem of photometric redshift estimation belongs to the regression task of data mining.
Thus the algorithms fit for regression can be applied for photometric redshift estimation. When
the sample is specified, a choice of approaches is needed. Comparison of different regressors depends on different regression metrics, such as the residual between the spectroscopic and photometric redshifts, $\Delta z=z_{\rm spec}-z_{\rm photo}$, and the mean absolute error $\sigma$. Another metric to determine the goodness of photometric redshift estimation is the fraction of test samples that satisfy $\mid\bigtriangleup z\mid=|z_{i}-\widehat{z_{i}}|<e$ (\citealt{sch17} and references therein).

The definition of mean absolute error $\sigma$ is as follows:
\begin{equation}
\sigma = \frac{1}{n}\sum_{i=0}^{n-1}|z_{i}-\widehat{z}_{i}|\\
\end{equation}
where $z_{i}$ is the true redshift, $\widehat{z_{i}}$ is the the predicted redshift value and $n$ is the sample size.

The fraction of test samples that satisfies $\mid\bigtriangleup z\mid=|z_{i}-\widehat{z_{i}}|<e$ is usually used to evaluate the redshift estimation, where $e$ is a given residual threshold,
\begin{equation}
f_{\mid\bigtriangleup z\mid<e}=\frac{N(|z_{i}-\widehat{z_{i}}|<e)}{N_{\rm total}}\\
\end{equation}

The typical values of $e$ are 0.1, 0.2 and 0.3. However, the redshift normalized residuals are often adopted,
\begin{equation}
\delta_{e}=\frac{N(|z_{i}-\widehat{z_{i}}|<e(1+z_{i}))}{N_{\rm total}}\\
\end{equation}

Once the methods have been chosen, the next important task is to determine an algorithm's hyperparameters which indicate how the machine learns. For $k$NN, the model parameter is only $k$ when taking Euclidean distance as metric measure and KD Tree as index. But for RF, main
parameters contain the maximum depth of individual
trees (max depth) and the number of trees in the forest
($n$ estimators). The goal is to find the optimal model parameters which optimize
the algorithm's performance. In reality we don't know these values in advance. Therefore a grid search is performed with
$K$-fold cross-validation, which means that the training sample is
split into $K$ subsamples, such that one subsample is left to
estimate the algorithm's performance while the remaining
subsamples are utilized to train the algorithm and construct the classifier/regressor. This process is done
$K$ times and finally the average performance is kept.
The entire process is repeated for every combination of
hyperparameters in the grid space and values that
optimize the performance are output. The grid for $k$NN has $k$ values from 10 to 30. The grid for RF has the hyperparameters: $n$\_$estimators$=$[50,100,200,300]$ and max$\_$depth=$[15,20,25]$ (12 combinations).

\subsection{Photometric redshifts estimation with one sample}

For the SDSS sample and the SDSS-WISE sample, the two samples are randomly divided into two parts: two-thirds as training sets and one-third as test sets. All of the model constructions are performed by 10-fold
cross-validation on the full training sets while the other test sets are applied to test the regressors ($k$NN and RF). Their performance, optimal model parameters and running time of the two
algorithms for model construction and predicting photometric redshifts of quasars are written
in Table~1. A comparison of photometric redshift estimation with spectroscopic redshifts by different methods is displayed in Figure~2. As listed in Table~1, for the SDSS sample with $k$NN, the percents ($\delta_{0.1}$, $\delta_{0.2}$ and $\delta_{0.3}$) in different $\frac{|\Delta z|}{1+z_{i}}$ intervals and the mean absolute error $\sigma$ are 62.53\%, 80.13\%, 87.17\% and 0.3326, respectively; for the SDSS-WISE sample with $k$NN, $\delta_{0.1}$, $\delta_{0.2}$ and $\delta_{0.3}$ and $\sigma$ are 79.40\%, 91.37\%, 95.28\% and 0.1931, separately; for the SDSS sample with RF, they are respectively 63.34\%, 80.48\%, 87.34\% and 0.3271; for the SDSS-WISE sample with RF, they are respectively 79.87\%, 91.37\%, 95.23\% and 0.1907.
For $k$NN, the running time of model construction and prediction is 242s with SDSS sample and 420s with the SDSS-WISE sample; while for RF, the running time is 37 628s and 36 762s, respectively. No matter for $k$NN or for RF, the accuracy of photometric redshift estimation improve apparently with both optical and infrared information compared to only the optical information. For the SDSS sample, the performance of RF is a little superior to that of $k$NN. Meanwhile, for the SDSS-WISE sample, the performance of RF is better than that of $k$NN except for $\delta_{0.2}$ and $\delta_{0.3}$, although their accuracy is comparable. If only considering speed, $k$NN shows its superiority.

\begin{figure}[!!!h]
\centering
\includegraphics[width=6cm]{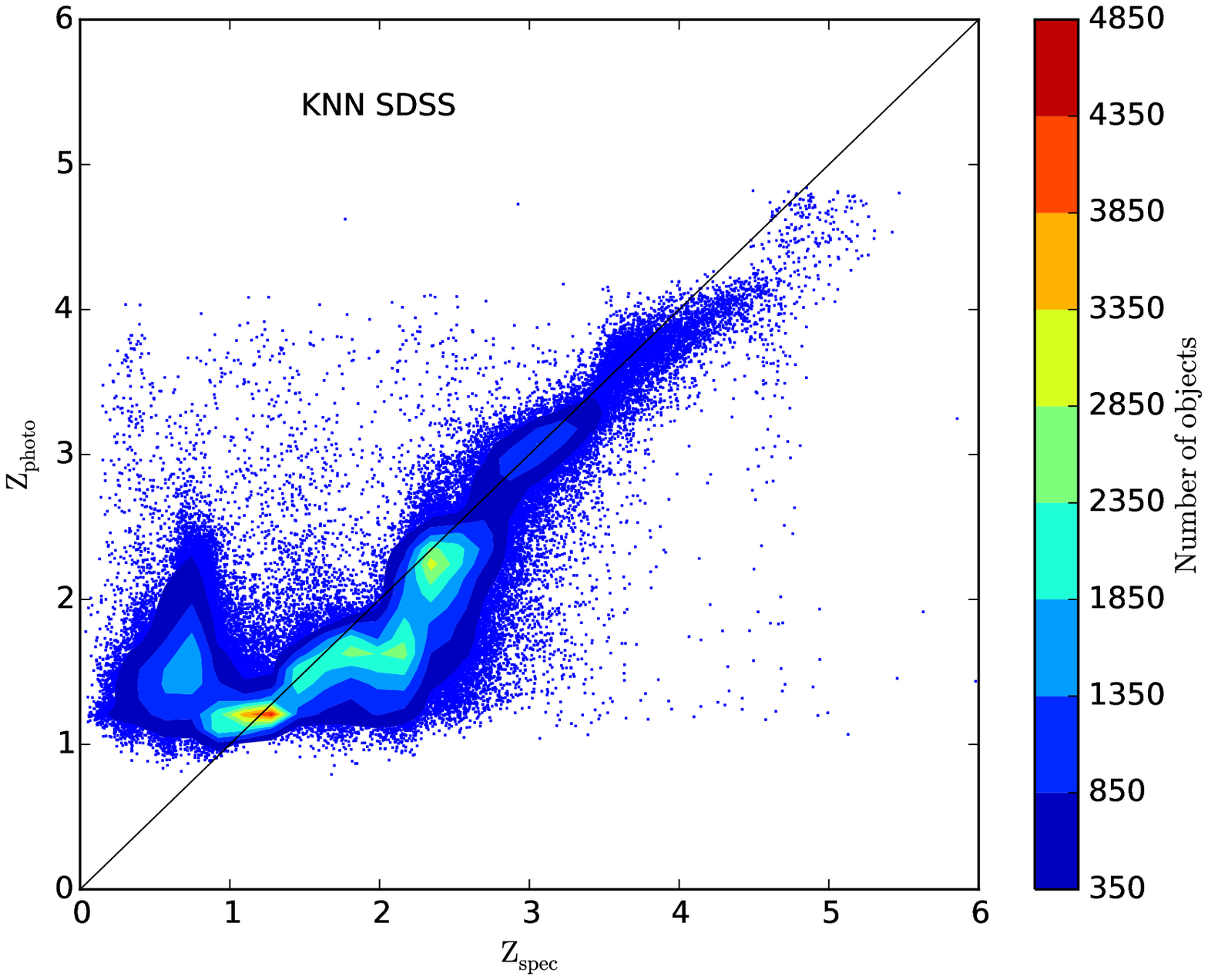}
\includegraphics[width=6cm]{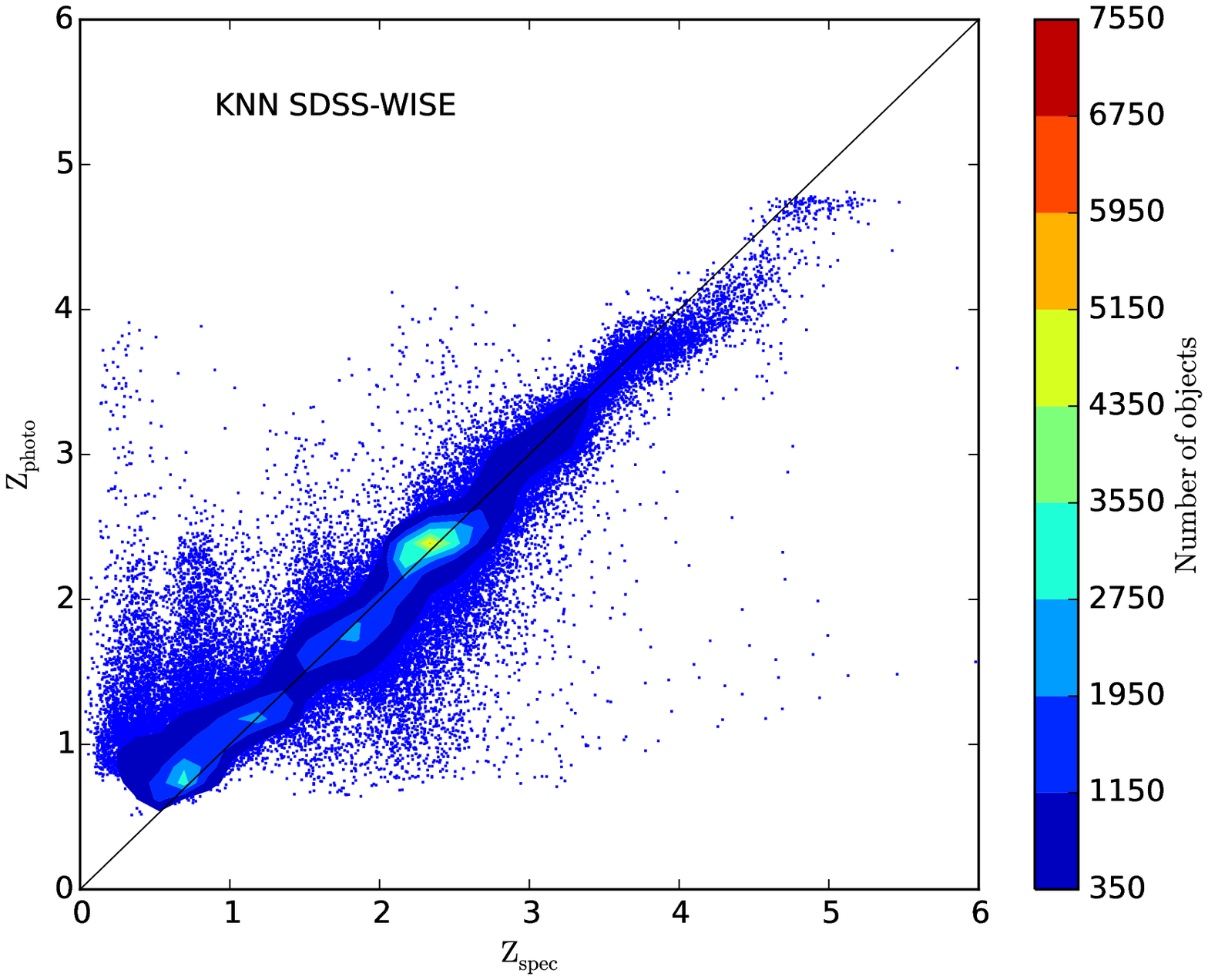}
\includegraphics[width=6cm]{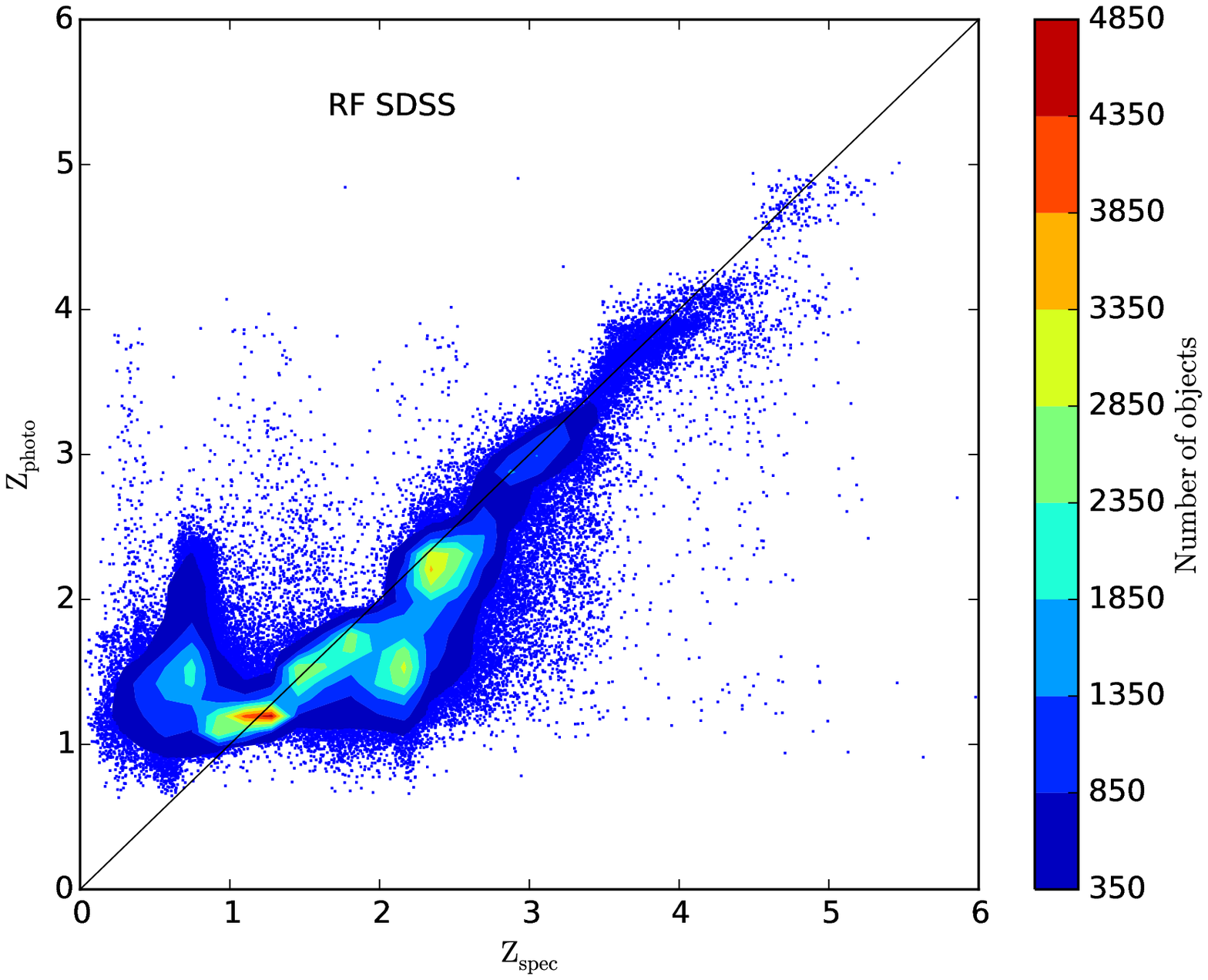}
\includegraphics[width=6cm]{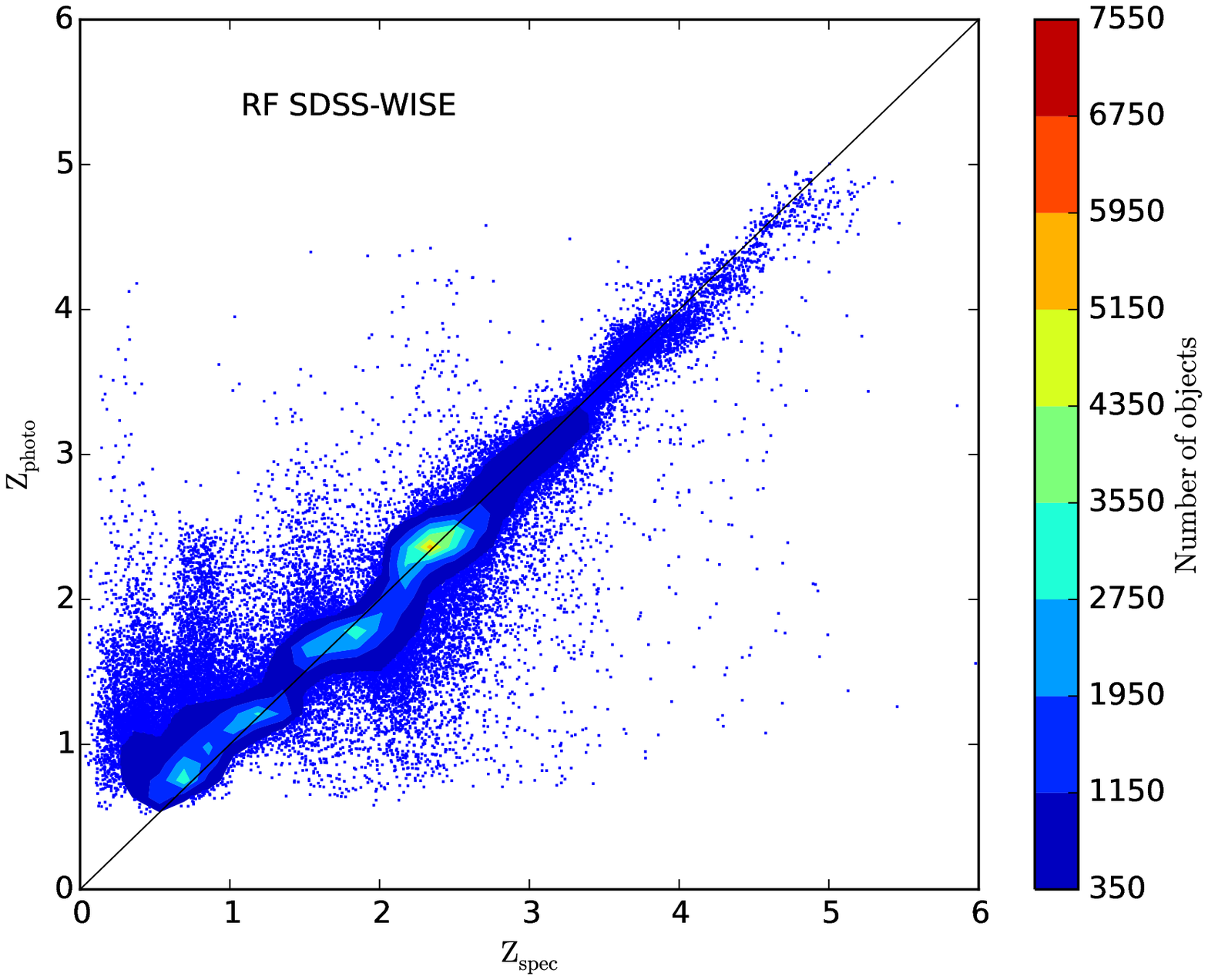}
\caption{Predicted photometric redshifts vs. spectroscopic redshifts. The color bars signify the number of objects per rectangular bin. The upper-left panel is based on SDSS test sample by $k$NN; the upper right panel is based on the SDSS-WISE test sample by $k$NN; the lower-left panel is based on the SDSS test sample by RF; the lower-right panel based on the SDSS-WISE test sample by RF.}
\label{fig2}
\end{figure}

\begin{table}[!!!h]
\begin{center}
\caption{Performance of photometric redshift estimation for different datasets with $k$NN and RF.
}\label{tab:perf} {
\small{
\begin{tabular}{lccccccccc}
\hline \hline
Data Set&Algorithm&Model Parameters&$\delta_{0.1}$(\%)  & $\delta_{0.2}$(\%) & $\delta_{0.3}$(\%)  & $\sigma$& Time(s) \\
\hline
SDSS&$k$NN&$k$=30&62.53&80.13&87.17&0.3326&242\\
SDSS-WISE&$k$NN&$k$=30&79.40&91.37&95.28&0.1931&420\\

\hline
SDSS&RF&$n$\_$estimators$=300&63.34&80.48&87.34&0.3271&37628\\
    &  &$max$\_$depth$=15    &     &     &     &        &\\
SDSS-WISE&RF&$n$\_$estimators$=300&79.87&91.37&95.23&0.1907&36762 \\
    &  &$max$\_$depth$=20    &     &     &     &        &\\
\hline
\end{tabular}}}
\end{center}
\end{table}

\subsection{Photometric redshift estimation with two subsamples}
Considering the redshift distribution due to the physical properties of quasars, quasars may be separated into different groups. To improve the performance of photometric redshift estimation, we put forward a scheme of first classification and second regression for photometric redshift estimation, specifically any new source is classified by a classifier in advance and subsequently its photometric redshift is predicted by a regressor.

For the detailed steps of photometric redshift estimation with two subsamples, see Figure~3. First, the quasar samples of SDSS and SDSS-WISE are divided into two subsamples: one with $0<z\le 2.2$ and the other with $2.2<z$. For the two subsamples, they are randomly segmented into two parts: two-thirds for training (A with redshift from 0 to 2.2 and B with redshift from 2.2 to 6) and one-third for testing (T1 with redshift from 0 to 2.2 and T2 with redshift from 2.2 to 6). With the training sets A and B, the classifier is created by 10-fold cross-validation. The testing sets T1 and T2 are applied as inputs for the classifier and then classified as T1A and T2A with redshift from 0 to 2.2 and T1B and T2B with redshift from 2.2 to 6. Second, the samples A and B are used as training sets to train regressors and represented as regressor\_A and regressor\_B, respectively. The testingsamples T1A and T2A are tested by regressor\_A while the testing samples T1B and T2B are tested by regressor\_B. Finally, the predicted results are obtained as result\_A and result\_B.

\begin{figure}[!!!h]
\centering
\includegraphics[width=15cm]{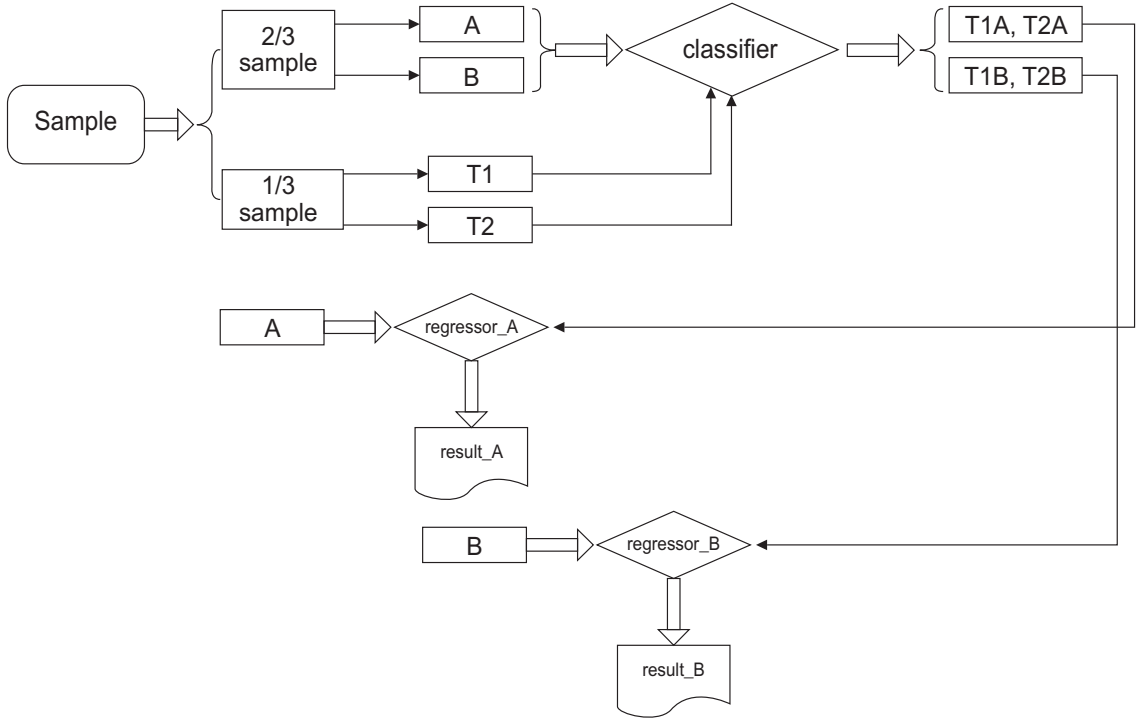}
\caption{Flow chart of photometric redshift estimation based on two subsamples.}
\label{fig3}
\end{figure}

In brief, the core algorithm of the classifier is RF, while the regressors adopt $k$NN and RF. For the SDSS and SDSS-WISE samples, they are randomly segmented into two-thirds for training and one-third for testing. The RF classifier is constructed by 10-fold validation with the full training set. The regressors ($k$NN and RF) are also built by 10-fold validation with the full training set. For convenience, RF is utilized for classification and $k$NN is for regression, RF\_KNN for short; RF is employed both for classification and regression, RF\_RF for short. For the SDSS sample with RF\_KNN, the optimal parameters of the RF classifier are
$n\_$estimators$=100$ and max$\_$depth$=15$; while for the SDSS-WISE sample with RF\_KNN, $n\_$estimators$=300$ and max$\_$depth$=25$. For the SDSS sample with RF\_RF, the optimal parameters of the RF classifier are
$n\_$estimators$=300$ and max$\_$depth$=15$; while for the SDSS-WISE sample with RF\_RF, $n\_$estimators$=300$ and max$\_$depth$=20$. For different subgroups, the performance of photometric redshift estimation for the SDSS and SDSS-WISE samples with $k$NN after classifying one sample into two subsamples by RF is indicated in Table~2, while the performance with RF is shown in Table~3. Comparison of photometric redshift estimation with spectroscopic redshifts by different methods is indicated in Figure~4. No matter if implementing RF\_KNN or RF\_RF, adding the infrared information is helpful to improve the accuracy of photometric redshift estimation, and the performance based on T2 is better than that based on T1. Comparing the results in Table~2 with those in Table~3, it is found that the performance of RF\_KNN is a little inferior to that of RF\_RF except for $\delta_{0.3}$ of SDSS T1 and SDSS-WISE T1, $\delta_{0.3}$ of SDSS-WISE T1+T2 and three $\delta$ values of SDSS-WISE T2. Considering the entire test sets (SDSS T1+T2 and SDSS-WISE T1+T2), RF\_RF manifests slightly better performance than RF\_KNN.

\begin{table}[!!!h]
\begin{center}
\caption{Performance of photometric redshift estimation for different datasets with $k$NN after classifying one sample into two subsamples by RF.
}\label{tab:perf} {
\small{
\begin{tabular}{lccccccccc}
\hline \hline
Data Set (Test set)&Algorithm&Model Parameters&$\delta_{0.1}$(\%)  & $\delta_{0.2}$(\%) & $\delta_{0.3}$(\%)  & $\sigma$\\
\hline
SDSS (T1)&RF\_$k$NN&$k$=30&54.07&71.79&84.44&0.3574\\
SDSS (T2)&RF\_$k$NN&$k$=30&84.58&89.31&90.12&0.2856\\
\hline
SDSS (T1+T2)&&&67.11&79.28&86.87&0.3267\\
\hline
\hline
SDSS-WISE (T1)&RF\_$k$NN&$k$=30&74.80&89.37&94.73&0.2037\\
SDSS-WISE (T2)&RF\_$k$NN&$k$=20&93.05&96.40&97.00&0.1710\\
\hline
SDSS-WISE (T1+T2)&&&80.97&91.75&95.50&0.1926\\
\hline
\end{tabular}}}
\end{center}
\end{table}

\begin{table}[!!!h]
\begin{center}
\caption{Performance of photometric redshift estimation for different datasets with RF after classifying one sample into two subsamples by RF.
}\label{tab:perf} {
\small{
\begin{tabular}{lccccccccc}
\hline \hline
Data Set (Test set)&Algorithm&Model Parameters&$\delta_{0.1}$(\%)  & $\delta_{0.2}$(\%) & $\delta_{0.3}$(\%)  & $\sigma$\\
\hline
SDSS (T1)&RF\_RF&$n$\_$estimators$=300&55.08&72.07&84.36&0.3550\\
    &  &$max$\_$depth$=15    &     &     &     &        \\
SDSS (T2)&RF\_RF&$n$\_$estimators$=300&84.77&89.55&90.31&0.2810\\
    &  &$max$\_$depth$=15    &     &     &     &        \\
\hline
SDSS (T1+T2)&&&67.74&79.52&86.90&0.3235\\
\hline
\hline
SDSS-WISE (T1)&RF\_RF&$n$\_$estimators$=300&75.77&89.55&94.52&0.2022\\
    &  &$max$\_$depth$=15    &     &     &     &        \\
SDSS-WISE (T2)&RF\_RF&$n$\_$estimators$=300&93.01&96.40&96.97&0.1660\\
    &  &$max$\_$depth$=20    &     &     &     &        \\
\hline
SDSS-WISE (T1+T2)&&&81.60&91.87&95.35&0.1900\\
\hline
\end{tabular}}}
\end{center}
\end{table}

\begin{figure}[!!!h]
\centering
\includegraphics[width=6cm]{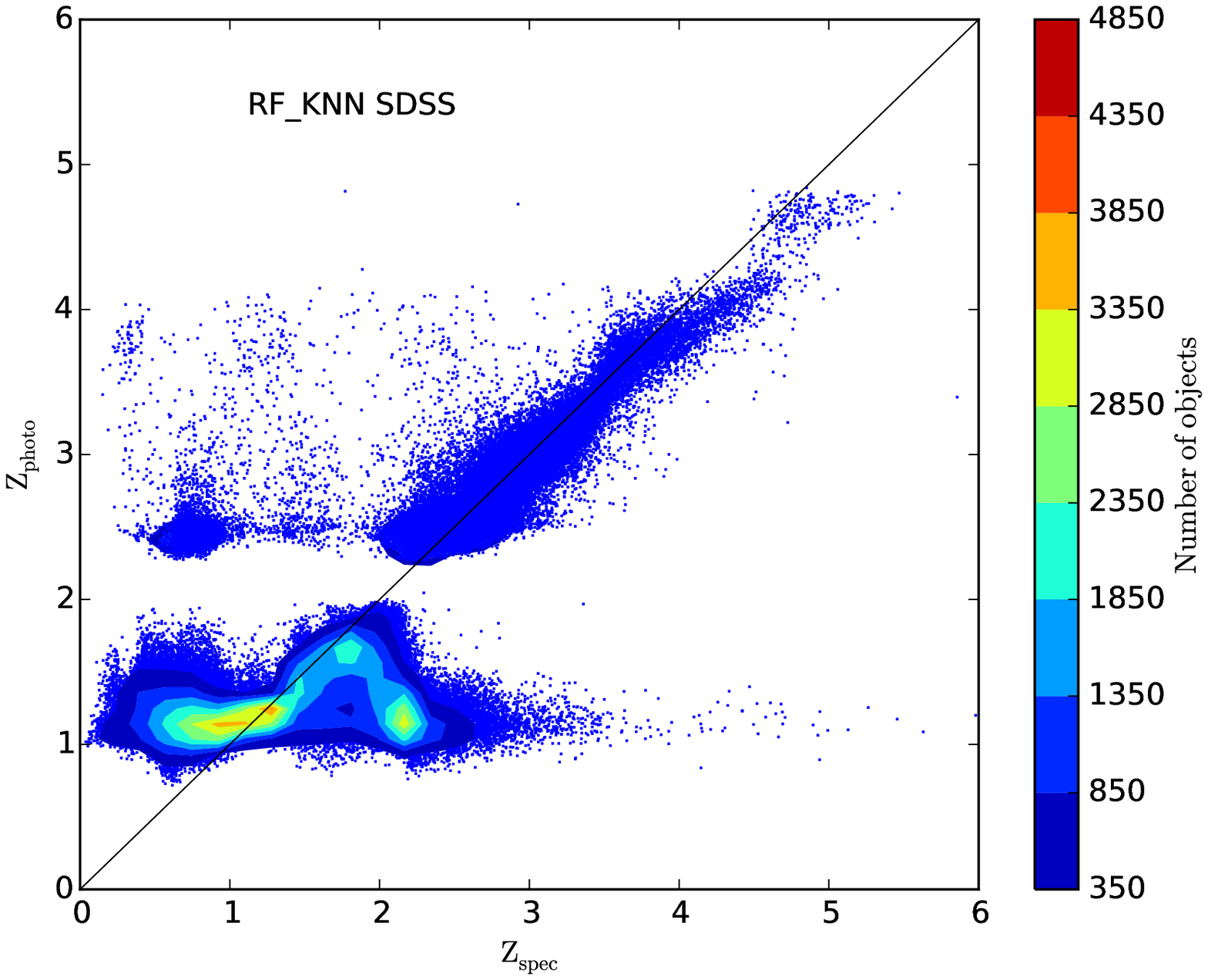}
\includegraphics[width=6cm]{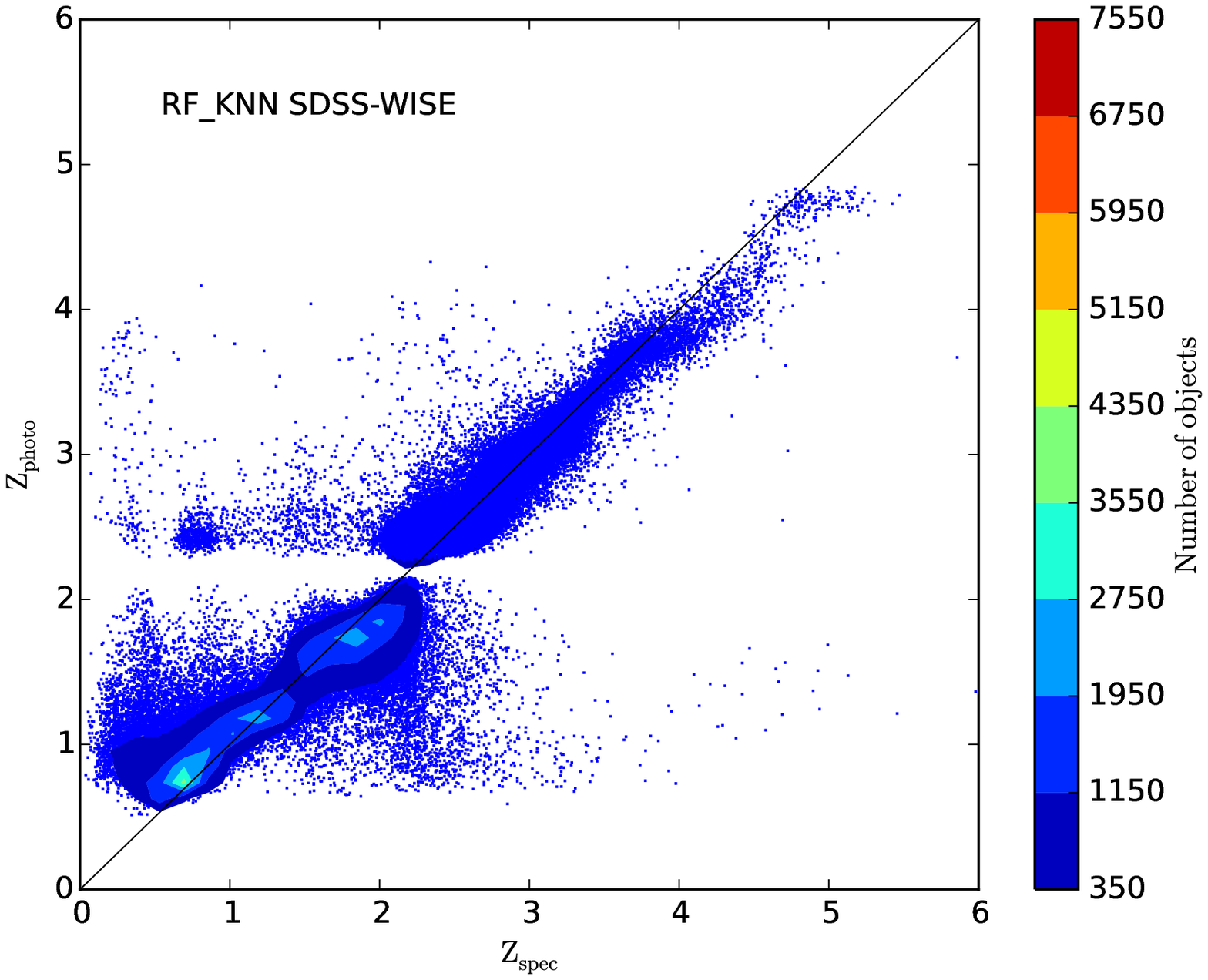}
\includegraphics[width=6cm]{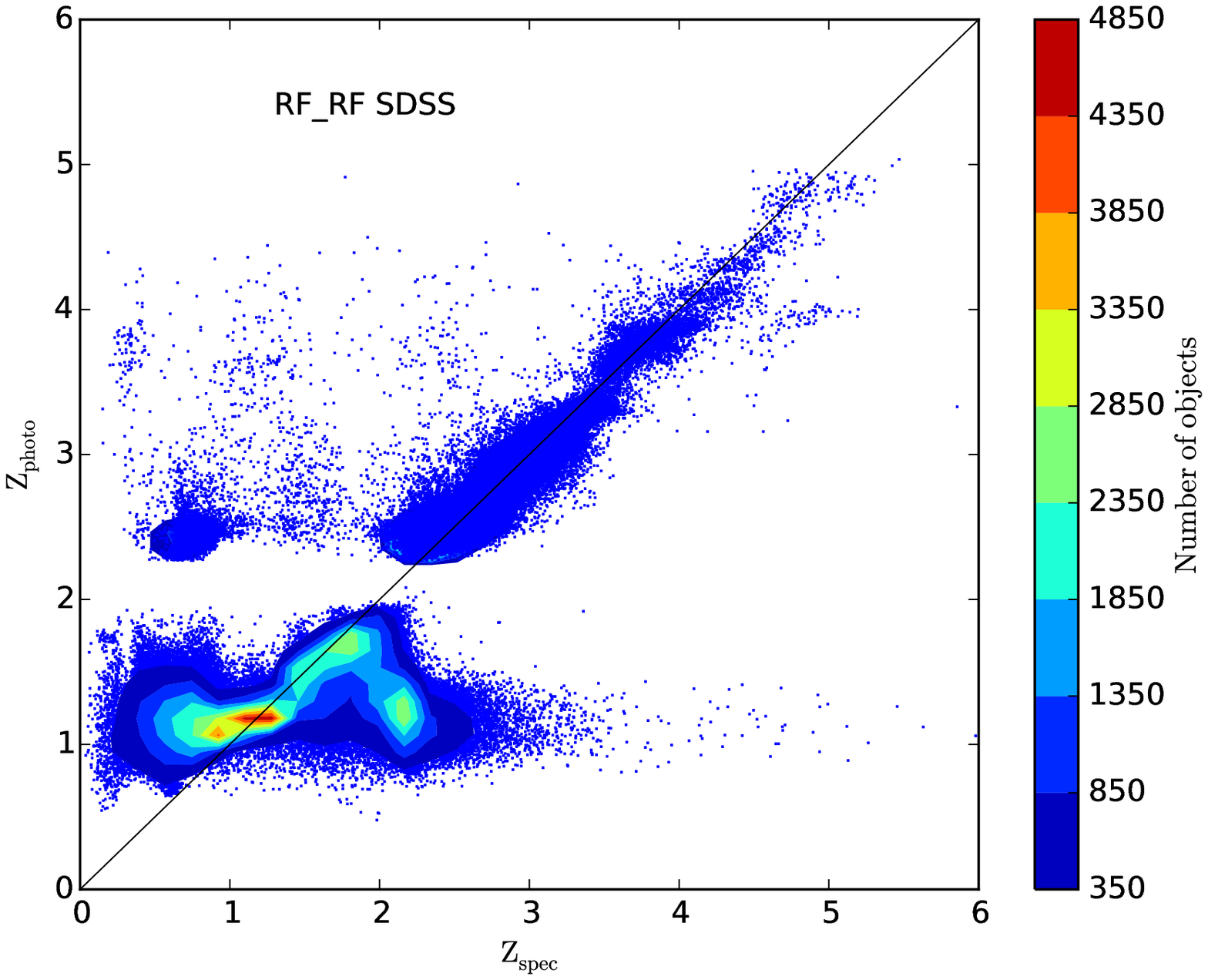}
\includegraphics[width=6cm]{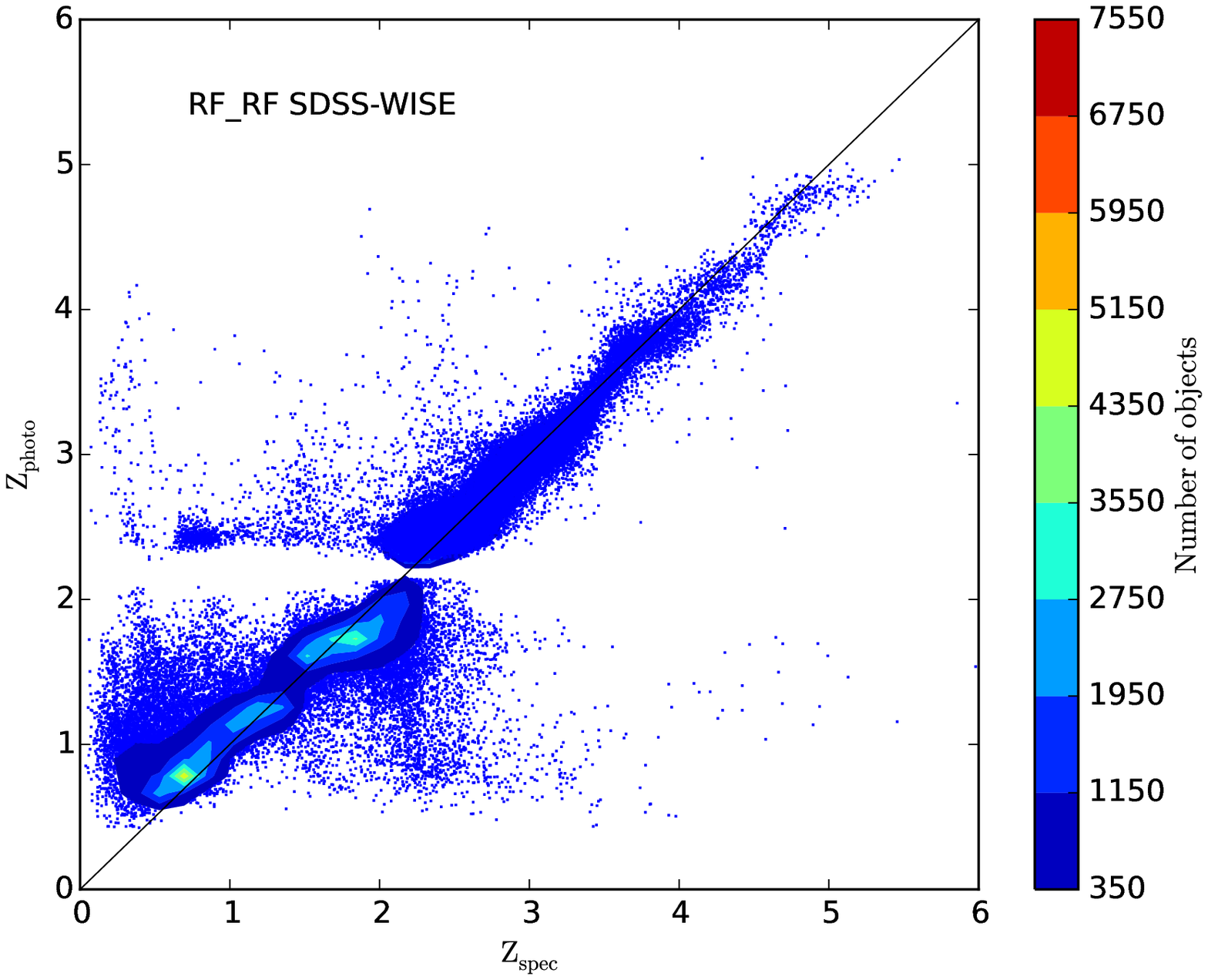}
\caption{Predicted photometric redshifts vs. spectroscopic redshifts for two subsamples. The color bars signify the number of objects per rectangular bin. The upper-left panel is based on the SDSS test sample by $k$NN; the upper-right panel is based on the SDSS-WISE test sample by $k$NN; the lower-left panel is based on the SDSS test sample by RF; the lower-right panel is based on the SDSS-WISE test sample by RF.}
\label{fig4}
\end{figure}

\subsection{Photometric redshift estimation with four subsamples}
Similar to Section 4.2, we put forward another scheme of first classification and second regression for photometric redshift estimation. To be different, the quasar samples of SDSS and SDSS-WISE are separated into four subsamples: ``vlowz" with $0<z \le1.5$, ``lowz" with $1.5<z\le 2.2$, ``midz" with $2.2<z\le 3.5$, and ``highz" with $3.5<z$. These four subsamples are randomly broken up into two parts: two-thirds for training ($a$ with redshift from 0 to 1.5, $b$ with redshift from 1.5 to 2.2, $c$ with redshift from 2.2 to 3.5, and $d$ with redshift from 3.5 to 6) and one-third for testing ($t1$ with redshift from 0 to 1.5, $t2$ with redshift from 1.5 to 2.2, $t3$ with redshift from 2.2 to 3.5 and $t4$ with redshift from 3.5 to 6). Based on training sets $a$, $b$, $c$ and $d$, the classifier is created. Then the classifier separates testing sets $t1$, $t2$, $t3$ and $t4$ into $t1a$, $t2a$, $t3a$ and $t4a$ with redshift from 0 to 1.5, $t1b$, $t2b$, $t3b$ and $t4b$ with redshift from 1.5 to 2.2, $t1a$, $t2c$, $t3c$ and $t4c$ with redshift from 2.2 to 3.5, and $t1d$, $t2d$, $t3d$ and $t4d$ with redshift from 3.5 to 6.0. Next, the samples $a$, $b$, $c$ and $d$ are used for training regressors, and four regressors are obtained, represented as regressor\_a, regressor\_b, regressor\_c and regressor\_d, respectively. The testing samples $t1a$, $t2a$, $t3a$ and $t4a$ are tested by regressor\_a, $t1b$, $t2b$, $t3b$ and $t4b$ by regressor\_b, $t1c$, $t2c$, $t3c$ and $t4c$ by regressor\_c, and $t1d$, $t2d$, $t3d$ and $t4d$ by regressor\_d. In the end, the predicted results are obtained as result\_a, result\_b, result\_c and result\_d, respectively. The detailed steps of photometric redshift estimation with four subsamples are shown in Figure~5.

\begin{figure}[!!!h]
\centering
\includegraphics[width=15cm]{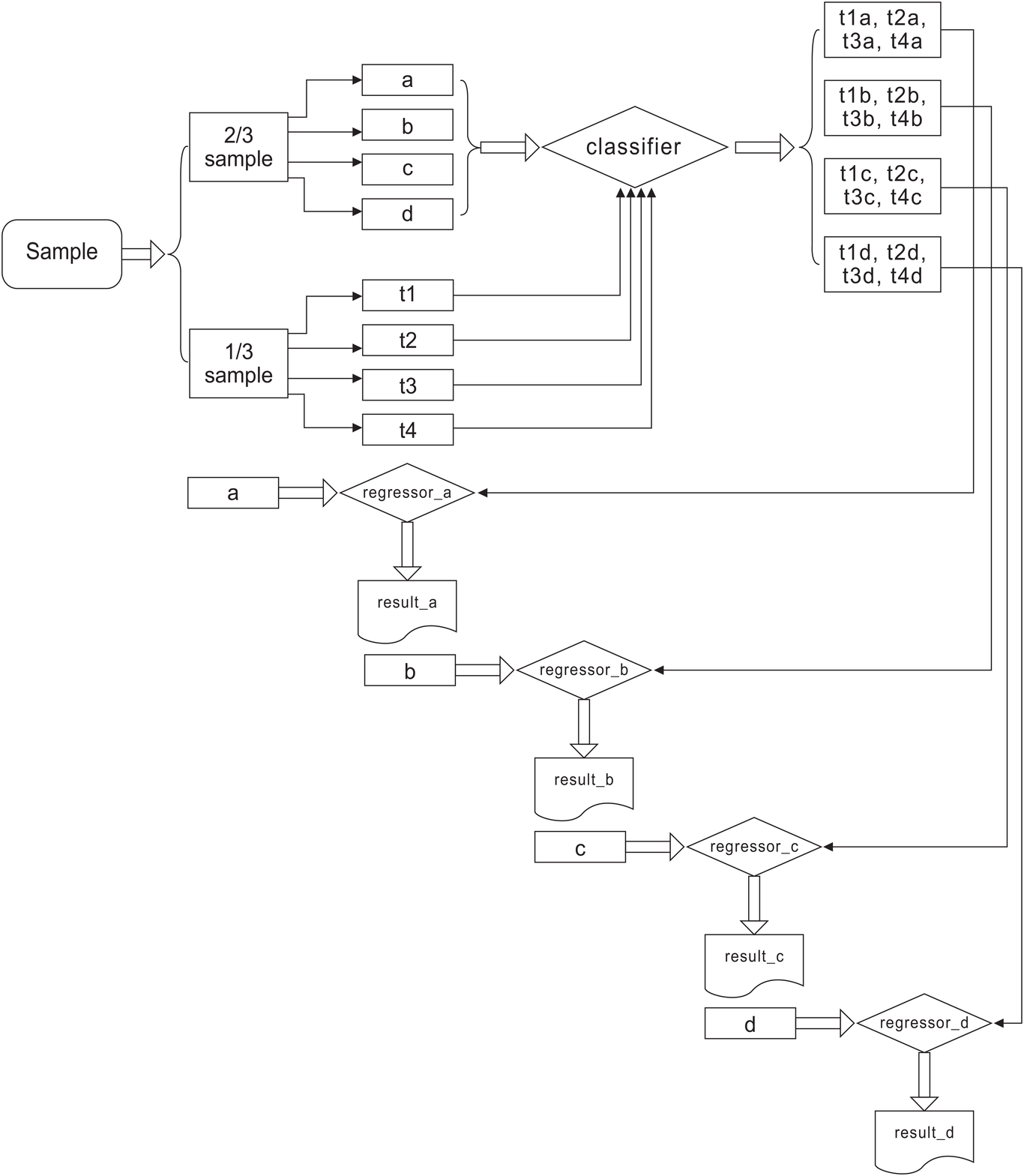}
\caption{Flow chart of photometric redshift estimation based on four subsamples.}
\label{fig5}
\end{figure}

In the whole process, RF is still adopted as the classification algorithm, while RF and $k$NN are utilized as the regression algorithms. The performance of photometric redshift estimation for the SDSS and SDSS-WISE samples with $k$NN after classifying one sample into four subsamples by RF is indicated in Table~4, while the performance with RF is shown in Table~5. Comparison of photometric redshift estimation with spectroscopic redshifts by different methods is indicated in Figure~6. For the SDSS sample with RF\_KNN, the optimal parameters of the RF classifier are
$n\_$estimators$=300$ and max$\_$depth$=15$; while for the SDSS-WISE sample with RF\_KNN, $n\_$estimators$=200$ and max$\_$depth$=20$. For the SDSS sample with RF\_RF, the optimal parameters of the RF classifier are
$n\_$estimators$=300$ and max$\_$depth$=15$; while for the SDSS-WISE sample with RF\_RF, $n\_$estimators$=300$ and max$\_$depth$=25$. To compare of the results in Table~4 with those in Table~5, given that there are only three $\delta$ values for the test sets (SDSS t4 and SDSS-WISE t4), the accuracy of RF\_KNN is better than RF\_RF, while only considering $\sigma$, RF\_RF is superior to RF\_KNN; given that there are only $\delta_{0.2}$ and $\delta_{0.3}$ for the test set SDSS t1, RF\_KNN shows better performance than RF\_RF while considering $\delta_{0.1}$ and $\sigma$, RF\_RF displays superiority over RF\_KNN; for the test set SDSS-WISE t1, RF\_RF is better than RF\_KNN except $\delta_{0.3}$. In terms of the entire test sets (SDSS t1+t2+t3+t4 and SDSS-WSIE t1+t2+t3+t4), RF\_RF achieves slightly better results than RF\_KNN.

\begin{table}[!!!h]
\begin{center}
\caption{Performance of photometric redshift estimation for different datasets with $k$NN after classifying one sample into four subsamples by RF.
}\label{tab:perf} {
\small{
\begin{tabular}{lccccccccc}
\hline \hline
Data Set (Test set)&Algorithm&Model Parameters&$\delta_{0.1}$(\%)  & $\delta_{0.2}$(\%) & $\delta_{0.3}$(\%)  & $\sigma$\\
\hline
SDSS (t1)&RF\_$k$NN&$k$=30&65.36&75.91&80.47&0.3719\\
SDSS (t2)&RF\_$k$NN&$k$=30&72.96&84.83&86.88&0.2905\\
SDSS (t3)&RF\_$k$NN&$k$=30&81.79&86.92&87.90&0.3181\\
SDSS (t4)&RF\_$k$NN&$k$=10&95.57&96.66&96.82&0.1948\\
\hline
SDSS (\small{t1+t2+t3+t4})&&&75.16&83.48&85.73&0.3235\\
\hline
\hline
SDSS-WISE (t1)&RF\_$k$NN&$k$=20&78.96&89.80&93.56&0.1949\\
SDSS-WISE (t2)&RF\_$k$NN&$k$=30&82.91&94.05&95.72&0.1885\\
SDSS-WISE (t3)&RF\_$k$NN&$k$=30&91.77&95.35&96.09&0.1833\\
SDSS-WISE (t4)&RF\_$k$NN&$k$=10&98.02&98.56&98.76&0.1498\\
\hline
SDSS-WISE (\small{t1+t2+t3+t4})&&&84.63&92.95&95.08&0.1885\\
\hline
\end{tabular}}}
\end{center}
\end{table}

\begin{table}[!!!h]
\begin{center}
\caption{Performance of photometric redshift estimation for different datasets with RF after classifying one sample into four subsamples by RF.
}\label{tab:perf} {
\small{
\begin{tabular}{lccccccccc}
\hline \hline
Data Set (Test set)&Algorithm&Model Parameters&$\delta_{0.1}$(\%)  & $\delta_{0.2}$(\%) & $\delta_{0.3}$(\%)  & $\sigma$\\
\hline
SDSS (t1)&RF\_RF&$n$\_$estimators$=200&65.65&75.88&80.27&0.3760\\
    &  &$max$\_$depth$=15    &     &     &     &        \\
SDSS (t2)&RF\_RF&$n$\_$estimators$=300&73.81&85.05&87.07&0.2854\\
    &  &$max$\_$depth$=15    &     &     &     &        \\
SDSS (t3)&RF\_RF&$n$\_$estimators$=300&82.04&87.16&88.16&0.3131\\
    &  &$max$\_$depth$=15    &     &     &     &        \\
SDSS (t4)&RF\_RF&$n$\_$estimators$=50&95.35&96.49&96.68&0.1935\\
    &  &$max$\_$depth$=15    &     &     &     &        \\
\hline
SDSS (\small{t1+t2+t3+t4})&&&75.56&83.62&85.82&0.3213\\
\hline
\hline
SDSS-WISE (t1)&RF\_RF&$n$\_$estimators$=300&80.43&90.16&93.35&0.1916\\
    &  &$max$\_$depth$=15    &     &     &     &        \\
SDSS-WISE (t2)&RF\_RF&$n$\_$estimators$=300&83.35&93.69&95.45&0.1860\\
    &  &$max$\_$depth$=15    &     &     &     &        \\
    SDSS-WISE (t3)&RF\_RF&$n$\_$estimators$=300&91.95&95.48&96.26&0.1770\\
    &  &$max$\_$depth$=15    &     &     &     &        \\
SDSS-WISE (t4)&RF\_RF&$n$\_$estimators$=200&97.31&98.30&98.52&0.1420\\
    &  &$max$\_$depth$=20    &     &     &     &        \\
\hline
SDSS-WISE (\small{t1+t2+t3+t4})&&&85.33&93.01&94.97&0.1843\\
\hline
\end{tabular}}}
\end{center}
\end{table}

\begin{figure}[!!!h]
\centering
\includegraphics[width=6cm]{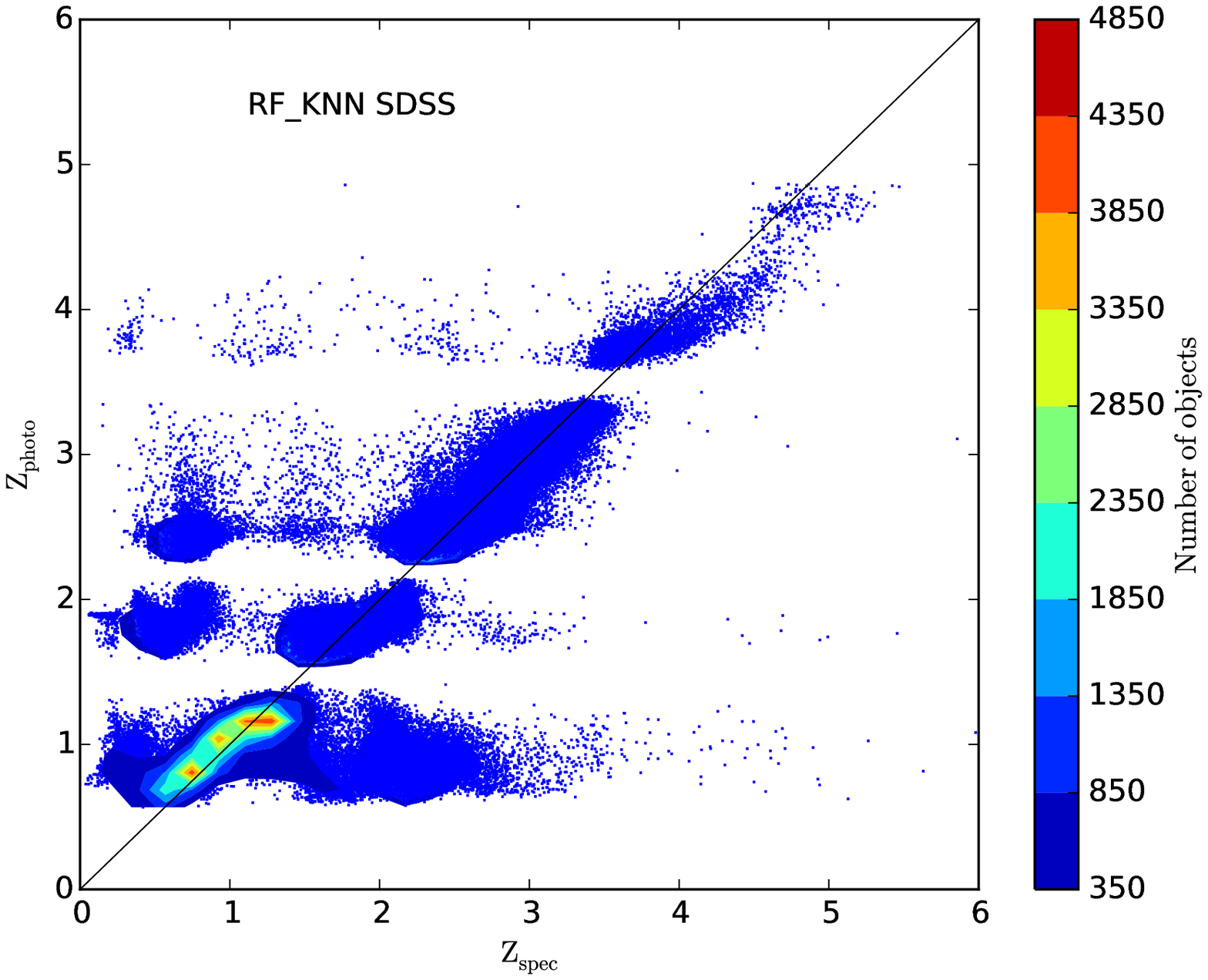}
\includegraphics[width=6cm]{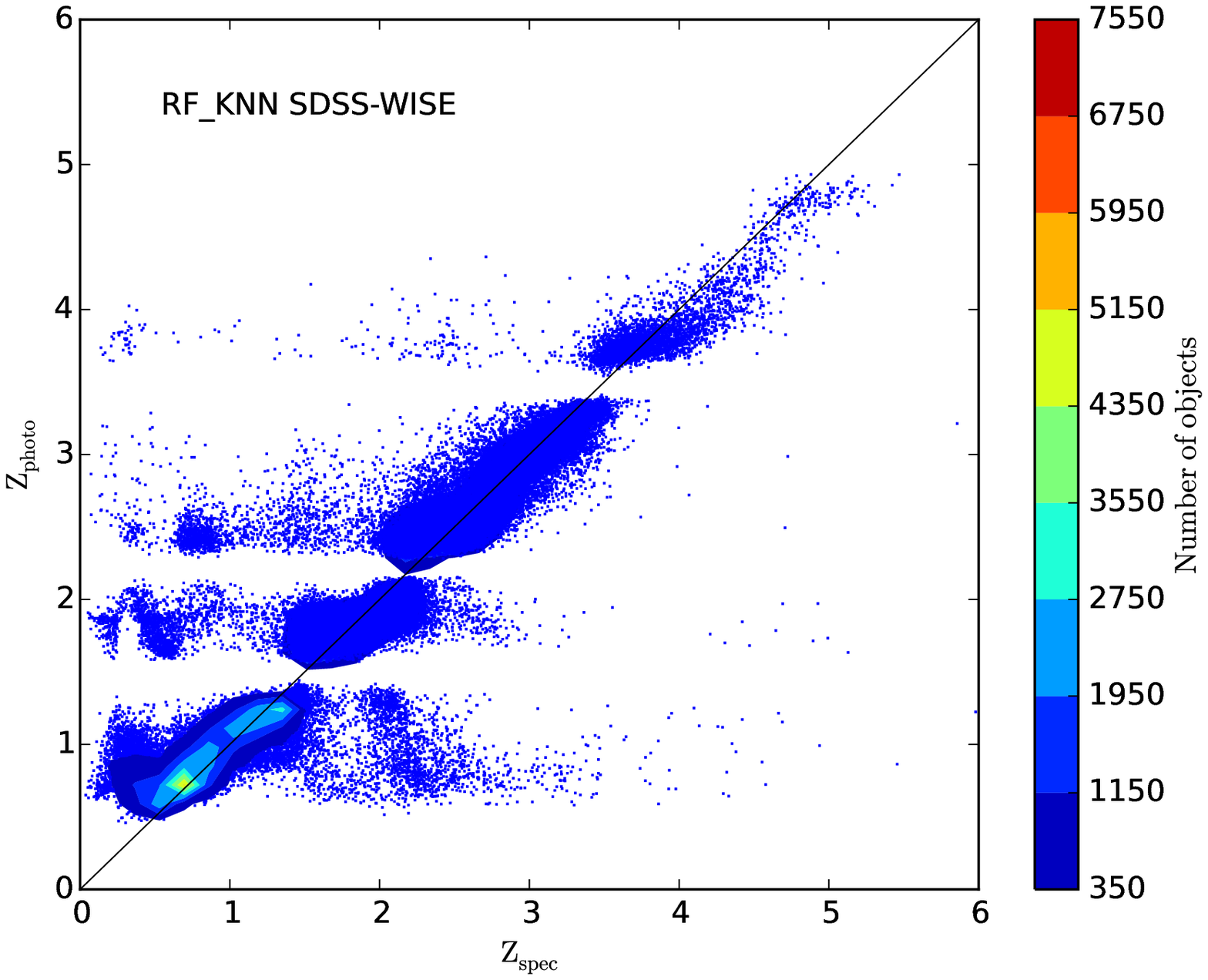}
\includegraphics[width=6cm]{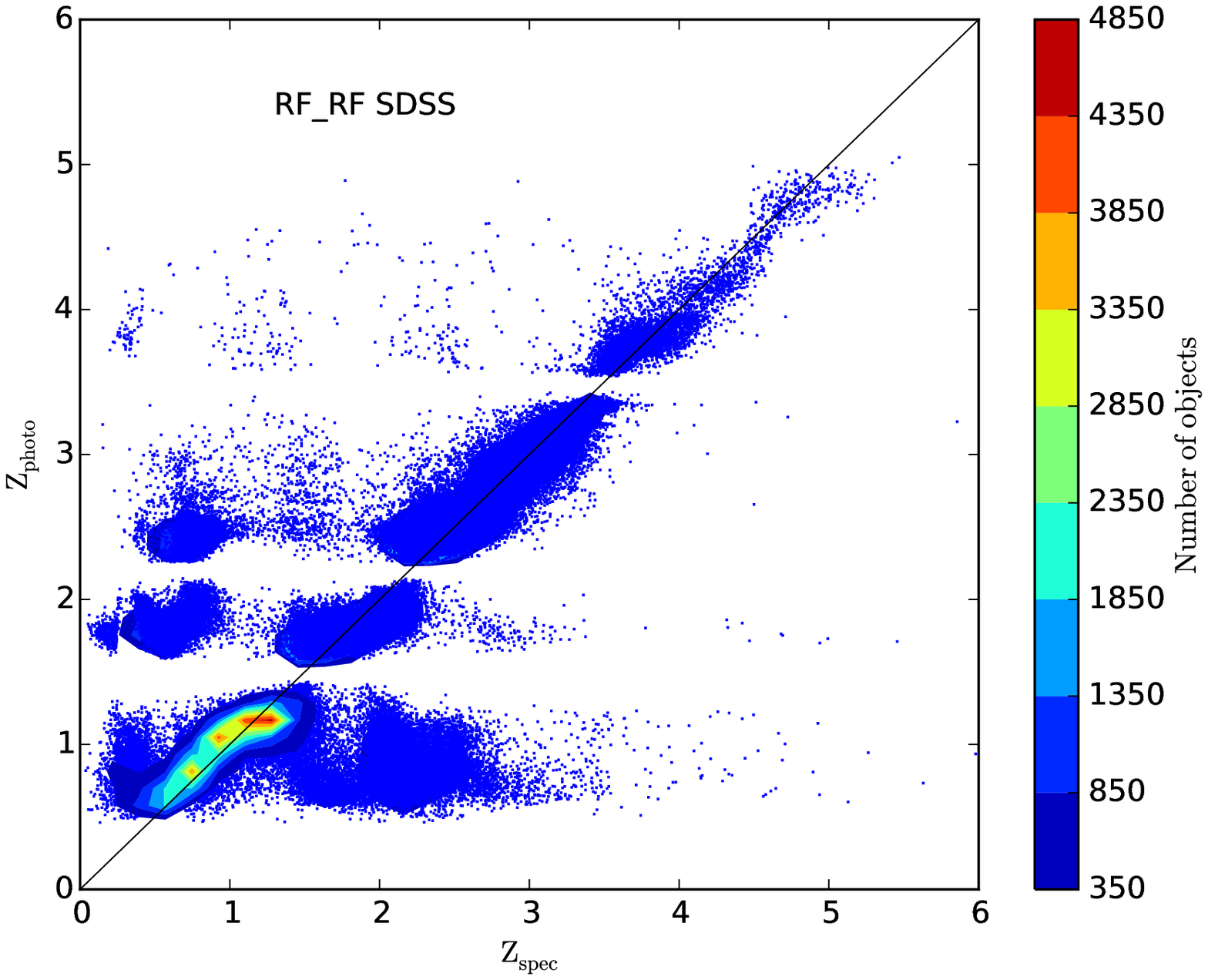}
\includegraphics[width=6cm]{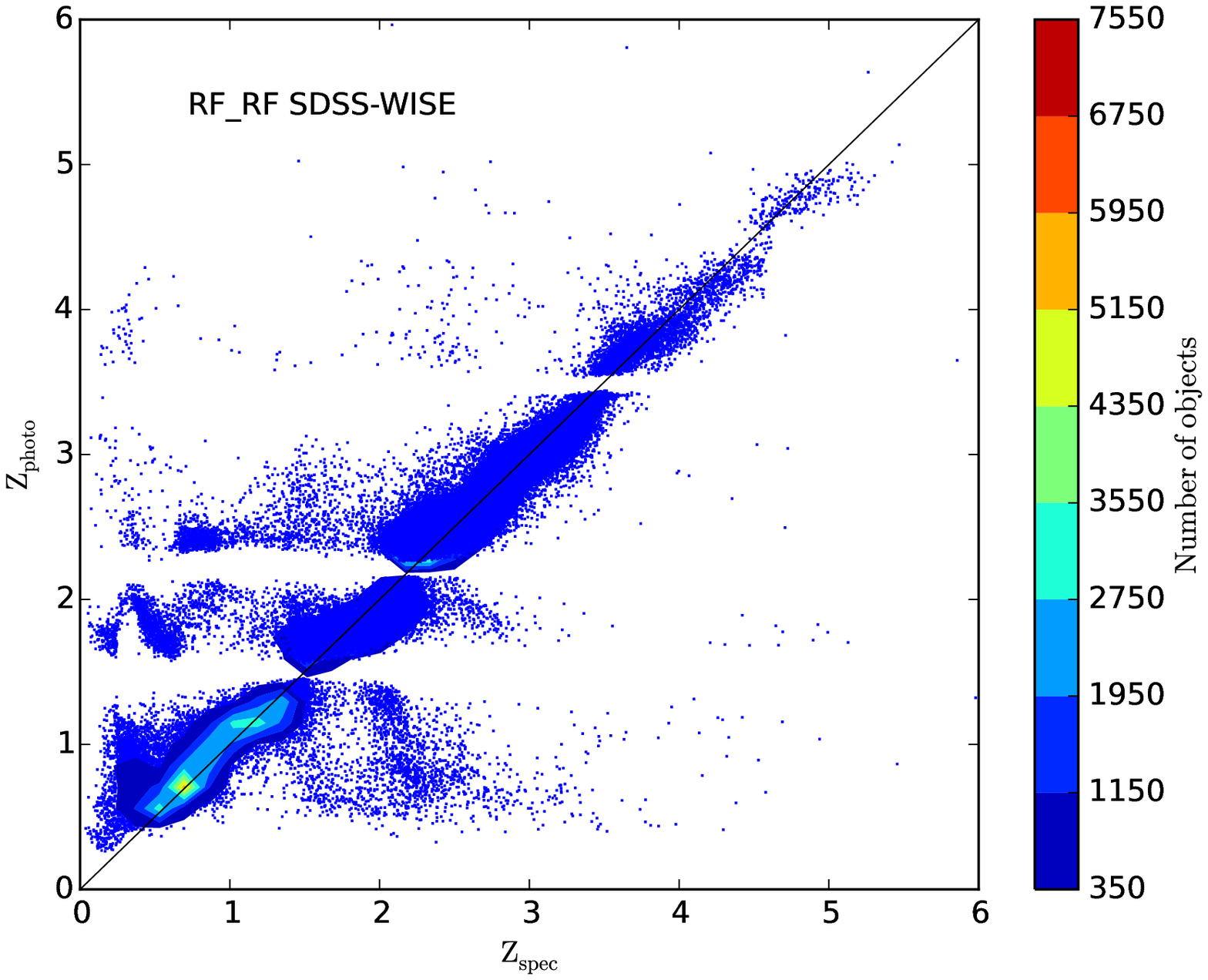}
\caption{Predicted photometric redshifts vs. spectroscopic redshifts for four subsamples. The color bars signify the number of objects per rectangular bin. The upper-left panel is based on the SDSS test sample by $k$NN; the upper right panel is based on the SDSS-WISE test sample by $k$NN; the lower-left panel is based on the SDSS test sample by RF; the lower right panel is based on the SDSS-WISE test sample by RF.}
\label{fig6}
\end{figure}

\section{Discussion}
The above results are summarized and compared in Table~6. The experimental results indicate that the accuracy of photometric redshift estimation can be generally improved by dividing the sample into subsamples and the accuracy of four subsamples is superior to that of two subsamples when not considering the percent in $\frac{|\Delta z|}{1+z_{i}}<0.3$; the performance with information both from optical and infrared bands enhances compared to that with only optical information; the four estimation metrics ($\delta_{0.1}$, $\delta_{0.2}$, $\delta_{0.3}$ and $\sigma$) all only improve with the SDSS-WISE sample divided into two subsamples. Therefore, the scheme of dividing the sample is indeed effective and the accuracy of four subsamples is better than that of two subsamples. It is evident that the accuracy is rather satisfying if accurately knowing the redshift range of new objects in advance. The accuracy further improves through the classification system and the improvement in accuracy depends on the accuracy of classification into subsamples. In reality, we do not know the redshift range of new objects beforehand. If we want to estimate better redshifts of new objects, then we need to judge their redshift range. Therefore, it is necessary to construct a classification system before estimating photometric redshifts. More information from more bands leads to performance of a classifier or a regressor becoming better. In addition, there is a lot of room for improvement from the perspective of percents in different redshift ranges since the percent in $\frac{|\Delta z|}{1+z_{i}}<0.3$ does not improve by the two schemes for most situations.

When the sample is classified into two/four subsamples, a discontinuity in the photometric redshift distribution exists due to misclassification near the cutoff. Because the accuracy of the classifier is much higher, the degree of discontinuity is much lower. Therefore, we may reduce the discontinuity by improving the accuracy of the classifier. In addition, when we utilize the photometric redshift catalog for further scientific study, we may adopt the estimated redshift value from two or four samples far from the cutoff, and keep the estimated value from one sample near the cutoff. Taking the SDSS-WISE sample into four subsamples by RF\_RF for example, we adopt the estimated redshift value from the regressor and keep the estimated value from one sample by RF near the three cutoff points ($\pm 0.3$), then the metrics ($\delta_{0.1}$, $\delta_{0.2}$, $\delta_{0.3}$ and $\sigma$) are 85.76\%, 93.28\%, 95.19\% and 0.1699, respectively. As a result, this method is applicable. To compare the performance of photometric redshift estimation, the true redshifts and estimated redsifts by different methods are depicted in Figure~7.

\begin{figure}[!!!h]
\centering
\includegraphics[bb=0 0 437 349,width=12cm]{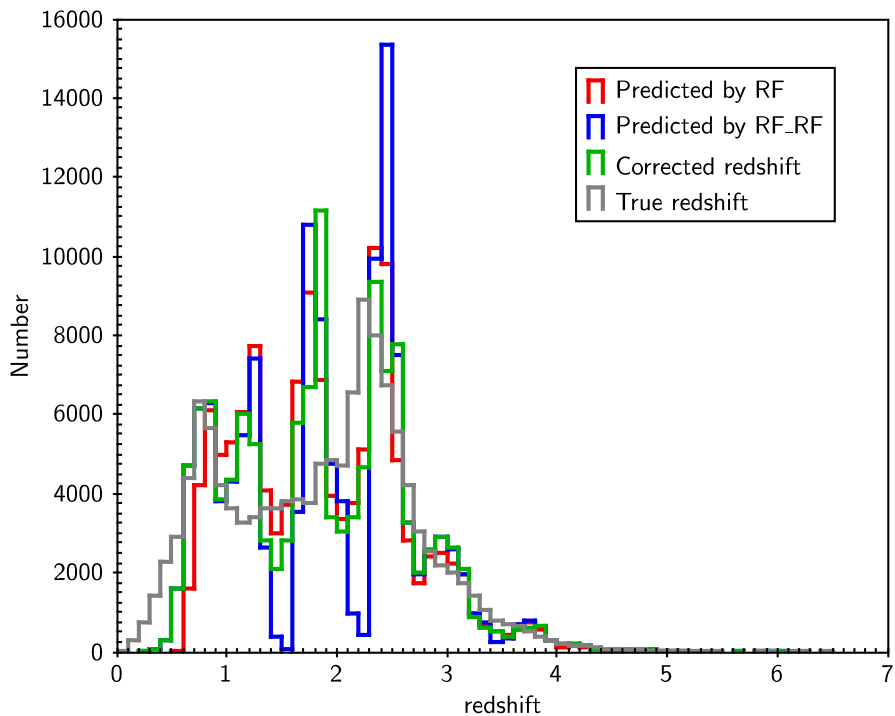}
\caption{Redshift distribution. Grey line represents true redshift; red line for estimated redshift from one sample by RF; blue line for estimated line from four samples by RF\_RF; green line for estimated redshift from four samples by RF\_RF and corrected near the cutoff.}
\label{fig6}
\end{figure}

In general, there are many factors influencing the accuracy of photometric redshift estimation, among which the adopted techniques and selected features are most important. However, other factors are also not neglected.
For example, \citet{sin11} presented the effects of including galaxy morphological parameters in photometric redshift estimation with an artificial neural network method. \citet{way11} found that the broad bandpass photometry of the SDSS in combination with precise knowledge of galaxy morphology was helpful to improve the accuracy of estimating photometric redshifts for galaxies.
\citet{soo18} studied the effects of incorporating galaxy morphology information in photometric redshift estimation, and found that the inclusion of quasar redshifts and associated object sizes in training improved the quality of photometric redshift catalogs and morphological information can mitigate biases and scatter due to bad photometry.
\citet{gom18} investigated improving photometric redshift estimation using GPZ by size information, post-processing and improved photometry.
All these factors may be considered in our strategy in future work.

Considering improving the robustness, flexibility and automation of approaches for photometric redshift estimation, various tools in this aspect are in development, such as IMPZ \citep{bab04}, EAZY \citep{bra08, bra10}, ArborZ \citep{ger10}, BPZ \citep{ben11}, Hyperz \citep{bol00, bol11}, LePHARE \citep{arn11}, ANNz \citep{col04, nie09, lah12}, PhotoZ \citep{sag13}, XDQSO \citep{bov13}, TPZ \citep{car13}, SOMz \citep{car14}, PhotoRApToR \citep{cav15a}, GAz \citep{hog15}, DAMEWARE \citep{cav15b}, TailZ \citep{gra16}, CuBANz \citep{sam16}, GPZ \citep{alm16b}, ANNz2 \citep{sad16} and Photo-z-SQL \citep{beck17a, beck17b}. \citet{abd11} compared six photometric redshift codes (ANNz, HyperZ, SDSS, LePHARE, BPZ and ZEBRA) for 1.5 million luminous red galaxies (LRGs) in SDSS DR6. Therefore algorithms turning into automated tools are of great value once they are successfully applied in a specified issue. This is important and necessary for astronomers with such convenient tools in the big data era \citep{zhang15}.

\begin{table}[!!!h]
\begin{center}
\caption{Performance of photometric redshift estimation with different datasets for different schemes.
}\label{tab:perf} {

\begin{tabular}{lccccccc}
\hline \hline
Data Set &Scheme&Algorithm&$\delta_{0.1}$(\%)  & $\delta_{0.2}$(\%) & $\delta_{0.3}$(\%)  & $\sigma$\\
\hline
SDSS &one sample&$k$NN& 62.53&80.13&87.17&0.3326\\
SDSS &two subsamples&RF\_$k$NN&67.11&79.28&86.87&0.3267 \\
SDSS &four subsamples&RF\_$k$NN&75.16&83.48&85.73&0.3235\\
\hline
SDSS-WISE &one sample&$k$NN&79.40&91.37&95.28&0.1931\\
SDSS-WISE &two subsamples&RF\_$k$NN&$80.97$&$91.75$&$95.50$&$0.1926$ \\
SDSS-WISE &four subsamples&RF\_$k$NN&84.63&92.95&95.08&0.1885\\
\hline
SDSS &one sample&RF& 63.34&80.48&87.34&0.3271\\
SDSS &two subsamples&RF\_RF&67.74&79.52&86.90&0.3235 \\
SDSS &four subsamples&RF\_RF&75.56&83.62&85.82&0.3213\\
\hline
SDSS-WISE &one sample&RF&79.87&91.37&95.23&0.1907\\
SDSS-WISE &two subsamples&RF\_RF&$81.60$&$91.87$&$95.35$&$0.1900$ \\
SDSS-WISE &four subsamples&RF\_RF&85.33&93.01&94.97&0.1843\\
\hline
\end{tabular}}
\end{center}
\end{table}

\section{Conclusions}
In general, the work on accuracy improvement of photometric redshift estimation of quasars focuses on algorithm choice and feature selection. We design two schemes of photometric redshift estimation by first classification and then regression, comparing the performance of RF and $k$NN with the SDSS and SDSS-WISE samples for the two schemes to the original scheme. We explore how to deal with the sample itself, and how the sample segmentation influences the estimation accuracy. Considering the experimental results, we are able to improve the estimation accuracy of photometric redshifts through first classification and then regression. In most of our experiments, the performance of dividing the sample into four subsamples is better than that of two subsamples with the two algorithms for the two samples, moreover the accuracy of both schemes improves compared to the original scheme, except for the percent in $\frac{|\Delta z|}{1+z_{i}}<0.3$. In addition, for the SDSS-WISE dataset, no matter for RF or $k$NN, all the four metrics of performance criterion improve based on the sample divided into two parts compared to the one sample or four subsamples (see Table~6). RF shows a little better performance than $k$NN but its speed is slower than $k$NN since $k$NN is based on KD-tree index. The accuracy with the SDSS-WISE sample is superior to that with the SDSS sample when the same method is adopted.
For the case of the SDSS-WISE sample divided into four subsamples by RF\_RF, the estimated redshifts are adopted from the regressor and the estimated redshifts by RF with one sample replace them near the three cutoff points ($\pm 0.3$), then the metrics ($\delta_{0.1}$, $\delta_{0.2}$, $\delta_{0.3}$ and $\sigma$) amount to 85.76\%, 93.28\%, 95.19\% and 0.1699, respectively. In other words, the strategy we put forward is effective. The accuracy of the classification system directly influences the performance of regression. The classification and regression also depend on the available information. As a result, information added from more bands is necessary to improve the accuracy of photometric redshift estimation and classification. In our next work, we will apply the databases (Pan-STARRS, future LSST, etc.) for this issue. Photometric redshift estimation of galaxies or other objects may be also improved by a similar strategy.

\begin{center} \bf Acknowledgment\end{center}

We are very grateful to the constructive suggestions of the referee. This paper is funded by the 973 Program 2014CB845700
and the National Natural Science Foundation of China under grants No.11873066 and
No.U1731109. We acknowledge the SDSS and WISE databases. Funding for the Sloan Digital Sky Survey (SDSS) IV has been provided by the Alfred P. Sloan Foundation, the U.S. Department of Energy Office of Science, and the Participating Institutions. SDSS-IV acknowledges support and resources from the Center for High-Performance Computing at the University of Utah. The SDSS web site is www.sdss.org. SDSS-IV is managed by the Astrophysical Research Consortium for the Participating Institutions of the SDSS Collaboration including the Brazilian Participation Group, the Carnegie Institution for Science, Carnegie Mellon University, the Chilean Participation Group, the French Participation Group, Harvard-Smithsonian Center for Astrophysics, Instituto de Astrof\'isica de Canarias, The Johns Hopkins University, Kavli Institute for the Physics and Mathematics of the Universe (IPMU) /University of Tokyo, Lawrence Berkeley National Laboratory, Leibniz Institut f\"ur Astrophysik Potsdam (AIP), Max-Planck-Institut f\"ur Astronomie (MPIA Heidelberg), Max-Planck-Institut f\"ur Astrophysik (MPA Garching), Max-Planck-Institut f\"ur Extraterrestrische Physik (MPE), National Astronomical Observatories of China, New Mexico State University, New York University, University of Notre Dame, Observat\'ario Nacional / MCTI, The Ohio State University, Pennsylvania State University, Shanghai Astronomical Observatory, United Kingdom Participation Group, Universidad Nacional Aut\'onoma de M\'exico, University of Arizona, University of Colorado Boulder, University of Oxford, University of Portsmouth, University of Utah, University of Virginia, University of Washington, University of Wisconsin, Vanderbilt University, and Yale University. This work makes use of data products from the Wide-field Infrared Survey Explorer (WISE), which is a joint project of the University of California, Los Angeles, and the Jet Propulsion Laboratory/California Institute of Technology, funded by the National Aeronautics and Space Administration.

\end{document}